\documentclass[11pt,twoside,letterpaper]{article} 
\usepackage{times,fancyhdr}
\usepackage[dvips]{graphicx}
\usepackage{amsmath,epsfig,graphicx,amssymb}
\usepackage{url}
\usepackage{cite}
\usepackage{hyperref}

\makeatletter 
\def\cleardoublepage{\clearpage\if@twoside \ifodd\c@page\else%
    \hbox{}%
    \thispagestyle{empty}%
    \newpage%
    \if@twocolumn\hbox{}\newpage\fi\fi\fi} 
\makeatother

\begin{document}
\title{Material science and impact crater formation
{\begin{flushleft}
\vskip 0.45in
{\normalsize\bfseries\textit{Chapter~6}}
\end{flushleft}
\vskip 0.45in
\bfseries\scshape }}
\author{\bfseries\itshape V.Celebonovic\thanks{E-mail address: vladan@ipb.ac.rs}\\
Inst.of Physics,Univ.of Belgrade,Pregrevica 118,11080 Belgrade,Serbia}
\date{January 11,2017.}
\maketitle
\thispagestyle{empty}
\setcounter{page}{251}
\thispagestyle{fancy}
\fancyhead{}
\fancyhead[L]{In: Horizons in World Physics,vol.291 \\ 
Editor: Albert Reimer, pp. {\thepage-\pageref{lastpage-}}} 

\fancyhead[R]{ISBN 978-1-53611-008-1  \\
\copyright~2017 Nova Science Publishers, Inc.}
\fancyfoot{}
\renewcommand{\headrulewidth}{0pt}

\begin{abstract} 
The surfaces of solid objects in our planetary system are dappled with craters. Some of them are due to impacts of various solid projectiles into the surfaces of the objects. A smaller part of these craters is of volcanic origin.
On the Earth,two most often mentioned such events are the "Tunguska event" of 1908. and the impact which led to the formation of the Barringer crater in Arizona. Impact craters are frequently analyzed within the "scaling theory", which is founded on dimensional analysis. The same problem can be treated by using standard laws of material science and condensed matter physics. In this chapter the two approaches will be compared and possibilities for future work indicated to some extent. Some preliminary conclusions concerning an impact into a granular target will be presented. 
\end{abstract}

\vspace{2in}

\noindent \textbf{PACS} 96.12.Kz,  96.15.Qr,  64.10.+h
\vspace{.08in} 

\noindent \textbf{Key Words}:impacts ; formation of craters ; scaling laws ; material science    
\vspace{.08in}

\newpage
\pagestyle{fancy}
\fancyhead{}
\fancyhead[EC]{V.Celebonovic}
\fancyhead[EL,OR]{\thepage}
\fancyhead[OC]{Material science and craters}
\fancyfoot{}
\renewcommand\headrulewidth{0.5pt} 
\section{Introduction}

The surfaces of most solid objects in the planetary system contain a certain number of craters. Some of these are of volcanic origin, while others are the results of impacts of small bodies into the target surfaces. The present paper is devoted to the impact craters. Their existence follows from the fact that a multitude of small solid bodies, remaining from the epoch of formation of the planetary system, is orbiting the Sun. Two important reasons motivate the studies of impact craters: analyzing these craters gives the opportunity of learning more about the projectiles which made them and therefore about the processes which have led to the formation of the planetary system. More important is the "applied" interest in these objects: the impact of a sufficiently large object into a sufficiently densely populated region on the Earth would provoke a catastrophe. A database of impact craters known on the surface of the Earth exists on the address:

\href{http://www.passc.net/EarthImpactDatabase/}{http://www.passc.net/EarthImpactDatabase/}

A historical event, often mentioned as an example of such impacts is the Tunguska event of June 30,1908. Another example of such an event is the impact which led to the creation of the Barringer crater in Arizona. Just a few years ago, an event similar to the one of 1908 occurred again in Russia near the city of Chelyabinsk (Artemieva and Shuvalov,2016). It seemed for decades that there was "no trace" of the object which provoked the Tunguska event. However, recent work (Anfinogenov et.al.,2014) seems to indicate that a piece of the original object has been found. Its velocity of landing was estimated as at least $v_{landing}=547$$m/s$. Chemically, it is a highly silicified gravelite sandstone, composed $98.5$ \% of $SiO_{2}$.

Developing the possibilities of predicting the place of an impact, the size of the possible crater, or the height of a possible tsunami, is extremely interesting and important for humanity. The importance of possible tsunamis occurring as consequences of impacts can be illustrated by a simple fact: namely, around $70$ \% of the surface of the Earth is covered by water. Accordingly, impact into water is more probable than impact into the solid part of the  surface of the Earth. For an illustrative example of a study of tsunamis and their consequences see, for example,(W$\ddot{u}$nnemann et al.,2010).
Laboratory facilities for studies of materials under shock compression exist in numerous institutions. A list of some of these facilities is available at the following address: 

\href{http://mygeologypage.ucdavis.edu/stewart/OLDSITE/ImpactLabs.html}{http://mygeologypage.ucdavis.edu/stewart/OLDSITE/ImpactLabs.html}.

This paper contains a comparison of the two main approaches to the problem of impacts and the resulting craters: the scaling theory (Holsapple 1993) and the approach based on standard laws of condensed matter physics ( Celebonovic 2013). Each of the following sections is devoted to brief outlines of one of these approaches, the section after to a comparison of their possibilities. 

An introductory account of a related problem is also presented. Namely, it was assumed in the calculations that the surface of the target is a crystalline material. What will happen if it is granular? The final part is devoted to the comparison of the two approaches. 

\section{The scaling theory}

What does the term "scaling" denote?. Scaling is defined as the application of some relation (called the {\it scaling law})  to predict the outcome of one event from the results of another. Parameters differing between the two events are called {\it scaled} variables. Another meaning of the term scaling is predicting the dependence of the outcome of a problem on its parameters ( Holsapple 1993). 

The form of scaling laws can be determined in three ways: by impact experiments, analytical calculations and approximate theoretical solutions. Because of the simplicity of their principle, one may be tempted to think that the impact experiments are very easy to perform. Projectiles of varied composition and mass are fired with different speeds into targets of differing chemical composition, and data are measured on the resulting craters.Such experiments are being discussed and performed for decades ( some examples are Press et al.,1960; Oberbeck 1971 ; Fujivara et al.,1977 ; Baldwin et al.,2007; Kadono et al.,2010; Suzuki et al.,2012) and they have given various interesting results. A common problem for most of these experiments is that the projectiles are launched with velocities below those of interest for studies of creation of large craters. In a similar kind of experiments, solid targets are shock compressed by the impact of short-lived laser beams. For a relatively recent report on a newly developed experimental platform for such experiments see, for example, (Gauthier, Fletcher, Ravasio et al.,2014).

Laws of physics needed for theoretical studies of impacts and the formation of craters are well known. These are the basic laws of classical physics,conservation of mass,momentum and energy. In order to introduce material parameters into these laws,one needs the knowledge of the equations of state ($EOS$) of the materials of both the target and the impactor. 
In laboratory experiments the chemical composition is of course well known. However, in applications to real astronomical problems, the chemical composition of possible impactors is known with a relative error which depends on a number of factors. 

Asteroids are observable only in reflected sunlight. The albedo depends on the chemical composition of an object, while for
  a given value of the albedo the apparent magnitude is a function of the distance of the object to the Earth. Therefore, an object made up of a material with a weak albedo, and which is far from the Earth, will have a weak reflection spectrum. As a consequence, its chemical composition will be determined with a large relative error.   
If the impactor is sufficiently massive and the speed of impact sufficiently high, in the moment of impact a transition solid$\rightarrow$ gas or solid $\rightarrow$ plasma occurs; the gas ( or plasma) cools rapidly, and the consequences of the impact can be analyzed within condensed matter physics. 

Approximate theoretical solutions are based on a simple idea: the initial phase of the problem is approximated as a "point source" of shock waves propagating throughout the target after the impact. The initial phase here designates the phase immediately after the impact. This approach was developed for studies of the effects of nuclear explosions. For details see (Holsapple 1993; Nellis, 2000) and references given there.
 
A good example of a scaling law is the problem of formation of a crater of volume $V$, as a consequence of the impact of an impactor of radius $r$, speed $v$ and mass density $\rho_{1}$ into a target (planet) having surface gravity $g$, material strength $X$, and mass density $\rho$. Material strength is loosely defined as the ability of a material to withstand load without failure. All material properties can be expressed as combinations of physical quantities having dimensions of stress and mass density. This implies ( Holsapple,1993) that the volume of an impact crater can be expressed as

\begin{equation}
V = f[\left\{r,v,\rho_{1}\right\},\left\{\rho,X\right\},g]
\end{equation}
where the first three variables describe the impactor, the following two the material making up the planet, and the surface gravity of the planet. This expression is completely general, and scaling models are derived from it by dimensional analysis. It follows from equation (1) that
\begin{equation}
\frac{\rho V}{m}=f_{1}[\frac{g r}{v^{2}},\frac{X}{\rho v^{2}},\frac{\rho}{\rho_{1}}] 
\end{equation}
where $m = \frac{4 \pi}{3} \rho_{1} r^{3} $ is the mass of the impactor. The quantity on the left side is the ratio of the mass of the material within the crater to the mass of the impactor.It is usually called cratering efficiency and denoted by $\pi_{V}$. The first term in the function is the ratio of the lithostatic pressure $\rho g r$ to the initial dynamic pressure $\rho v^{2}$ generated by the impactor. The lithostatic pressure at a certain depth is defined as the pressure exerted by the material above it. This ratio is denoted by $\pi_{2}$; the second term is the ratio of the material strength to the dynamical pressure, denoted by $\pi_{3}$. The final term is the ratio of the mass densities. 
If all the parameters of eq.(2) were known, or could be measured or calculated, it would not be a particular problem to determine the volume of an impact crater. Finding  the general solution of this equation is an open problem. As a consequence, solutions of this equation are usually studied in two limiting cases: the "strength" regime and the "gravity" regime. 
The "strength" regime is the situation in which the strength of the surface material is larger than the lithostatic pressure. Practically speaking, this implies impactors with diameters of approximately one meter.This means that 
\begin{equation}
\frac{\rho V}{m}=f_{1}[\frac{X}{\rho v^{2}}] 
\end{equation}
 where it was assumed that the ratio of the densities is approximately one. In this regime, the volume of the impact crater increases linearly with the volume of the impactor, its mass and its energy. Any dimension of the crater increases with the radius of the impactor. 
In the opposite case, when the diameter of the impactor is of the order of a kilometer or more, the lithostatic pressure is bigger than the material strength,meaning that 
\begin{equation}
\frac{\rho V}{m}=f_{1}[\frac{g r}{v^{2}}] 
\end{equation}
This is the definition of the "gravity" regime. 
Various experiments (discussed in Holsapple,1993) have been performed on the dependence of $\pi_{V}$ on $\pi_{2}$, the result being an exact power law. This can be explained, as discussed in (Holsapple,1993) by the assumption that whenever there is a dependence on the impactor size and speed, it is actually the dependence on its kinetic energy. This idea was used in the early sixties, in scaling from a nuclear event called "Teapot ESS" to the creation of the Meteor Crater in Arizona.

The idea that the consequences of an impact depend on the kinetic energy of the impactor is equivalent to the "point source" approximation. The kinetic energy is given by $\frac{1}{2} m v^{2}$. Taking the cube root, introducing the mass density, and dropping the numerical factor, one gets the function
\begin{equation}
C = r \rho^{1/3} v^{2/3}
\end{equation} 
which can be generalized to
\begin{equation}
C = r \rho^{\mu} v^{\nu}
\end{equation} 
Using this, equation (1) becomes
\begin{equation}
V = f[r \rho^{\mu} v^{\nu},\rho_{1},X,g]
\end{equation} 
It can be shown by dimensional analysis (Holsapple, 1993) that in the strength regime the volume of a crater is given by 
\begin{equation}
V \propto\frac{m}{\rho_{1}}\times(\frac{\rho_{1} v^{2}}{X})^{3\mu/2}\times (\frac{\rho}{\rho_{1}})^{1-3\nu}
\end{equation}
and a similar expression can be derived for the gravity regime. A way for determining values of scaling exponents is the analysis of data in impact cratering experiments (such as Suzuki, 2012). A possibility, important for astronomical and/or geophysical applications, is fitting this equation to measured values of various parameters in it. Once they are known for a given material (or materials) calculations referring to the formation of the impact craters become possible.
\section {Condensed matter physics}

Surfaces of objects in the solar system on which impact craters exist are solid. As the impactors are solid objects, the question is what (if anything) can be concluded about the impacts by using laws of condensed matter physics and all kinds of measurable parameters of the surfaces of the targets and (possibly) of the pieces of the impactors. These possibilities will be explored to some extent in the present section, using general laws of physics and results of the present author. 

In order to form a crater in the material of the target, a projectile must have some minimal impact velocity. This value was determined in (Celebonovic and Sochay,2010). The condition for the formation of a crater was defined in that paper as the equality of the kinetic energy of a unit volume of the material of the impactor with the internal energy of the unit volume of the material of the target. It was shown there that this speed is given by
\begin{equation}
v^{2} = \frac{\pi^{2}}{5 \rho_{1}} \frac{(k_{B} T)^{4}}{\hbar^{3}} (\frac{\partial P}{\partial \rho})^{- 3/2}
\end{equation}
where $\rho_{1}$ is the mass density of the impactor, $T$ the temperature of the target, and $P,\rho$ are the pressure and mass density of the material of the surface of the target. Clearly, the minimal value of the speed of the impactor in the moment of impact depends also on the equation of state of the material of the target. The dimensions of the impactor and of the resulting crater were not taken into account. As a test,this expression was applied to the case of an impactor made up entirely of olivine $(Mg,Fe)_{2}SiO_{4}$. It was shown that the minimal impact sped of such an object should be $16.3 km/s$. For comparison, note that the impact velocity of a real object, asteroid $99942$ Apophis, is estimated to be between $13$
and $20 km/s$. This means that two completely different methods: celestial mechanics and condensed matter physics give mutually close results on the same problem.

 $99942$ Apophis is an interesting object for such a comparison, because celestial mechanics indicates that there exists a small but non-zero  probability that it collides with the Earth on April 13,2036. (Giorgini et.al.,2008). This possible date of impact is sufficiently far in the future, so there are strong chances that new observations will improve the orbit of Apophis and diminish the probability of impact. It is known (Delbo et.al.,2007) that the diameter of this asteroid is $270\pm60$ m. This may seem small, but the impact of an object of this size into the surface of the Earth would certainly have serious consequences.

The final result of any impact is a crater. If the kinetic energy of the impactor is high enough, and if the target has a suitable value of the heat capacity, a consequence of the impact will be heating of the target. Depending on the kinetic energy of the impactor, the target may heat enough so as to melt, and possibly even evaporate at the point of impact. In this regime condensed matter physics obviously  cannot be applied. Regardless of the amount of heating in the impact, the outcome is always the same: a certain quantity of material of the target gets "pushed aside" at the point of impact, implying the creation a crater of certain dimensions. The aim of the calculations outlined here is to draw conclusions about the impactor using measurable dimensions of the crater and various parameters of the target. 

The problem of formation of impact craters was recently expressed as the following analogous problem in condensed matter physics: how big must be the kinetic energy of the impactor in order to produce a hole of given dimensions in a target material with known parameters (Celebonovic,2013)? It was assumed that the material of the target is a crystal, that one of the usual types of bonding exists in it, and that as a consequence of the impact the target does not melt, so that condensed matter physics can be applied. The problem of heating in impacts has recently been discussed (Celebonovic,2012; Celebonovic and Nikolic,2015). 

This calculation starts from a simple physical idea: the kinetic energy of the impactor must be greater than or equal to the internal energy of some volume, denoted by $V_{2}$, of the material of the target. 
The kinetic energy of the impactor of mass $m_{1}$ and speed $v_{1}$ is obviously
\begin{equation}
E_{k} = \frac{1}{2} m_{1} v_{1}^{2}
\end{equation}
and the internal energy $E_{I}$ consists of three components: the cohesion energy $E_{C}$, the thermal energy $E_{T}$ and $E_{H}(T)$ - the energy required for increasing the temperature of the material at the point of impact by an amount $\Delta T$. Therefore,
\begin{equation}
E_{I} = E_{C}+E_{T} + E_{H}(T)
\end{equation} 
and the condition for the formation of an impact crater as a consequence of an impact is 
\begin{equation}
E_{I} = E_{k}
\end{equation}   
Expressions for various terms in $E_{I}$ exist in standard literature. Details of the calculation are available in (Celebonovic,2013) and the final result for the energy condition which must be satisfied to enable the formation of an impact crater is given by 
\begin{eqnarray}
3 k_{B} T_{1} N\nu [1-\frac{3}{8}\frac{T_{D}}{T_{1}}-\frac{1}{20}(\frac{T_{D}}{T})^{2}+\frac{1}{10}(\frac{T_{D}^{2}}{T T_{1}})\nonumber\\
+(\frac{1}{560})(\frac{T_{D}}{T})^{4}-\frac{1}{420}\frac{T^{4}}{T^{3}T_{1}}-\frac{3 \bar{u}^{2}\rho\Omega_{m}}{n p \nu k_{B} T_{1}}]
= \frac{2 \pi\rho_{1}}{3}r_{1}^{3}v_{1}^{2}
\end{eqnarray}
The number $N$ is equal to the ratio of the volume of the crater, and the volume of the elementary crystal cell,$v_{e}$: $N=V/v_{e}$. The meaning of other symbols is: $k_{B}$ Boltzmann's constant,$T$ the initial temperature of the target,$T_{1}$ the temperature to which the target heats, $T_{D}$ the Debye temperature of the target,$\rho_{1}$,$r_{1}$ $v_{1}$ - mass density,radius and impact velocity of the projectile,$p$,$n$ - parameters of the inter-atomic interaction potential in the material of the target,$\nu$ the number of particles in the elementary crystal cell,$\bar{u}$ the speed of sound in the material of the target and $\Omega_{m}$ is the volume per particle pair. 

Equation (13) may at first sight look as being very complicated. It fact, it is simply an expression of the law of conservation of energy. Its main result is that it links parameters of the impactor, with those of the material of the target, which was the aim of the calculation. Therefore, it opens up the possibility of drawing conclusions on the impactor by using measurable parameters of the target material. 

This expression was applied to a well known case - the Barringer crater in Arizona, for which most of the experimental parameters are known. Assuming that the material of the crater is pure Forsterite ($Mg_{2}SiO_{4}$ ), and making plausible assumptions about the other parameters of eq.(13), it was obtained that $v_{1}\cong 41 km/s$, which is far larger than existing estimates obtained by using celestial mechanics. 

A possible solution of this problem is to assume that the material of the target is a chemical mixture. Suppose that only 10 percent of the material is Forsterite, and keep all the other parameters constant. This will give the value of $v_{1}\cong 15 km/s$, for the impact speed, which is much closer to results of celestial mechanics. Details of this calculation are available in (Celebonovic,2013) . 

The calculation outlined above was performed using the notion of cohesive energy of solids. The problem is that the cohesive energy is a very "impractcal"quantity: it is defined as the energy needed to transform a sample of a solid into a gas of widely separated atoms (Marder,2010). As a consequence of this definition, it is difficult to measure experimentally and it is not related to the strength of solids measurable in experiments. 
A much more "practical" notion is the stress. It is defined as the ratio of the force applied on a body to the cross section of the surface of a body normal to the direction of the force. After an impact,a crater will form if stress in the material becomes sufficiently high for the formation of a fracture in the material of the target. 

The critical value of the stress needed for the occurrence of a fracture in a material is given by (Tiley, 2004)
\begin{equation}
\sigma_{C} = \frac{1}{2} \left(\frac{E\chi\tau}{r_{0} w}\right)^{1/2}
\end{equation}     
where $E$ is Young's modulus of the material, $\chi$ is the surface energy,$\tau$ is the radius of curvature of the crack, $r_{0}$ the inter-atomic distance at which the stress becomes zero and $w$ is the length of a crack which preexists in the material. Defined in this way, $\sigma_{C}$ has the dimensions of pressure.

In the moment of impact, the kinetic energy of the impactor is used for fracturing and heating the material of the target. Therefore:
\begin{equation}
E_{k} = \sigma_{C} V + C_{V} V (T_{1}-T_{0})
\end{equation}
where $V$ is the volume of the crater formed as a result of the impact,$C_{V}$ is the heat capacity of the target material and $T_{0}$ the initial temperature of the target. At this point, one encounters the problem of finding a suitable mathematical approximation of the shape of a crater, in order to be able to make an analytical estimate of the volume $V$.  In accordance with recent experiments (Suzuki et al.,2012) the volume of the crater is approximated by 
\begin{equation}
V = \frac{1}{3} \pi b^{2} c 
\end{equation}
where $b$ is the radius of the "opening" of the crater and $c$ denotes its depth. 

It will be assumed that the impactor is a sphere of radius $r_{1}$, made up of a material of density $\rho_{1}$ having impact velocity $v_{1}$. Its kinetic energy is given by $E_{k} = \frac{2\pi}{3} \rho_{1} r_{1}^{3} v_{1}^{2}$. It follows from eq.(15) that 
\begin{equation}
T_{1} = T_{0} + \frac{1}{C_{V}} (\frac{E_{k}}{V}-\sigma_{C})
\end{equation}  
and after some algebra (Celebonovic, 2014)
\begin{equation}
V = \frac{2\pi}{3} \frac{\rho_{1} r_{1}^{3} v_{1}^{2}}{\alpha C_{V} T_{0} + \sigma_{C}}
\end{equation}
where $T_{1}-T_{0} = \alpha T_{0}$, with $\alpha > 0$.
Equation (18) can be transformed into the following form
\begin{equation}
V = \frac{E_{k}}{\alpha C_{V} T_{0} + \sigma_{C}}
\end{equation}
This result shows that the crater volume is a linear function of the kinetic energy of the impactor. Turning to terrestrial experiments, raw data on impactor energies and the resulting crater volumes in (Suzuki et al.,2012), can be fitted by an equation of the form
\begin{equation}
V[m^{3}] = V_{0}\times\exp[E[J]/c]
\end{equation} 
 with $V_{0} = (4\pm2)\times 10^{- 7} m^{3}$ and $c = (583\pm56) J$.

For sufficiently low energies $E$ (or low values of the ratio $E/c$) it becomes possible to keep only the first order terms. Accordingly eq.(20) reduces to the form 
\begin{equation}
V-V_{0}\cong(V_{0}/c)E.
\end{equation}
Combining with results of the calculations reported here, it follows that 

$V_{0}/c = 1/(\alpha C_{V}T_{0}+\sigma_{C})$. 
The conclusion is that the results of the calculations reported here are relevant to low kinetic energies 
of the impactors.  

Calculations outlined here open up the possibility of making various estimates of physical quantities occurring in the equations. Using known experimental data, and taking that the most abundant mineral at the site of the Barringer crater is $SiO_{2}$ it was shown in (Celebonovic, 2014) that in the moment of impact the site heated up to $T_{1}\approx 1300 K$. For another terrestrial entity, the Kamil crater on the border of Egypt and Sudan, it was shown that $\sigma_{C}\cong1.56 \times10^{8} J/m^{3}$.
\section{A granular target?}

It was assumed so far in the present paper that the surface of the target can be approximated as a crystal. This approximation is often but not always physically realistic. For example, should an impact occur in the Sahara desert, the sand there could certainly {\bf not} be approximated as a crystal. In general terms, the question encountered here can be expressed in the following way: how do the consequences of an impact change if the material of the target is granular and not a crystal? From the point of view of condensed matter physics, this "transition" is extremely interesting. Granular matter physics at the end of the last century was at the level of condensed matter physics around $1930.$ (de Gennes,1999). 
We shall concentrate on two particular aspects: the shape of impact craters when formed in a granular material, as compared to their shape in a crystal, and changes in the quantity of energy needed to heat a granular material compared to the energy needed to heat the same amount of crystalline material for the same temperature difference.  

 Knowledge of the crater shape is needed for the determination of the volume of the crater. In the present paper, the volume of a crater was approximated by eq.(16), in accordance with experimental data of Suzuki et.al (2012). On the other hand, experiments with normal incidence of a solid spherical impactor into a deep non-cohesive granular bed, have shown that the profile of a crater can be fitted by

\begin{equation}
z=z_{c}+\sqrt{b^{2}+c^{2} r^{2}}
\end{equation} 
where $z_{c}$,$b$ and $c$ are fitting parameters (de Vet and de Bruyn,2007). Inserting this expression for the profile of a crater into eq.(16), and using the same notation in both cases, one gets the following result for the difference of crater volumes formed after an impact into a crystal ($V_{1}$) and granular material ($V_{2}$):
\begin{equation}
V_{1}-V_{2} = \frac{1}{3} \pi z r^{2} - \frac{1}{3} \pi r^2 [z_{c}+\sqrt{b^{2}+c^{2} r^{2}}]
\end{equation}
A simple calculation shows that this difference can become equal to zero for certain values of $z$,$c$ and $b$. Physically, this means that there exist some conditions of impacts which lead to craters of equal volume in crystal and granular targets,assuming that all the other conditions are the same in the two cases.

Another interesting question related to the "granularity" of the target is the heating of its material. This problem was analyzed in (Celebonovic,2012 ; Celebonovic and Nikolic,2015). However, in both of these papers it was assumed that the target is crystalline,which is the case often but not always.

Apart from terrestrial examples of granular materials,similar kind of targets was found in outer parts of the planetary system. It was shown back in 1998., that the asteroid $25143$ Itokawa consists of several pieces loosely bound together. Such asteroids later got the name "rubble pile" asteroids,and the question is how will such an object heat as a consequence of an impact.

Rubble piles are obviously {\it porous},and the the main physical parameter characterizing them is the porosity, defined as the following ratio
\begin{equation}
\phi=\frac{V_{V}}{V_{T}} < 1
\end{equation}

where $V_{V}$ denotes the  volume of voids and $V_{T}$ the total volume of the object. The value of $V_{T}$ can be decomposed as follows:
\begin{equation}
V_{T} = V_{1}+V_{V} = V_{1}+\phi V_{T}
\end{equation}
which means that
\begin{equation}
V_{T} = \frac{V_{1}}{1-\phi}
\end{equation} 
where $V_{1}$ denotes the volume of the "solid component" of $V_{T}$. 

The quantity of energy needed to heat a volume $V_{T}$ of a material having the specific heat $C_{V}$ by a temperature difference of $\Delta$T is given by
\begin{equation}
Q=C_{V} V_{T} \Delta T = (C_{V1}+C_{V2}) \times \frac{V_{1}}{1-\phi} \times \Delta T
\end{equation}
where $C_{V1}$ denotes the specific heat of the "solid component" of $V_{T}$ and $C_{V2}$ is the specific heat of the voids. Finally,
\begin{equation}
Q = \frac{C_{V1}+C_{V2}}{1-\phi} V_{1} \Delta T
\end{equation}
which is the expression for the quantity of energy needed to heat up for $\Delta$ T a volume $V_{1}$ of a solid having the specific heat $C_{V1}$, the porosity $\phi$ and the specific heat of voids $C_{V2}$. 

The value of the specific heat $C_{V1}$ depends on the chemical composition of the material. If the composition is known, then this value can either be measured or calculated. More interesting is the problem of the specific heat of the voids,denoted here by $C_{V2}$. The unknown quantity here is the composition of the voids. If they contain only the vacuum,and if there is no source of thermal radiation within the voids, the specific heat $C_{V2}$ will be zero. However, if the voids are filled with some gas,then the value of $Q$ will depend on the ratio of the two specific heats. 

How does this expression compare with the result for a pure solid? Simple reasoning shows that
\begin{equation}
Q = Q_{S} + \frac{C_{V2} V_{1}}{1-\phi} \Delta T
\end{equation}
 where $Q_{S}$ is the quantity of energy needed to heat a "pure" solid. Clearly, this result strongly depends on the values of $C_{V2}$ and the porosity $\phi$. The implication is that some more energy is needed to heat up a porous than  a non-porous material,with all the parameters being the same. Obviously, the closer the value of $\phi$ is to $1$, the bigger the value of $Q$ will be. 

\section {Comparing the two approaches}

In this contribution we have outlined to some extent two approaches to the problem of the impact craters on the surfaces of solid bodies in the planetary system. One is the so called scaling theory and the other is standard condensed matter physics. Both approaches have a similar goal: using available experimental data, and 
relevant laws of physics, draw as much conclusions as possible on the impacts and the impactors. 

Scaling theory aims at linking the craters with "celestial" origin with those resulting from man made,classical or nuclear explosions. Scaling in such a way gives encouraging results. This approach is very general, which is excellent, but there exists the problem of treatment of  phase transitions   However,the main method of work with scaling theories is dimensional analysis. One of the results of the scaling approach is that the volume of a crater formed after an impact depends also on the mass density of the target. The same conclusion can be reached within the condensed matter physics approach (Celebonovic, 2014).

The approach based on condensed matter physics is based on well known physical laws. However, by its very nature,this approach has an inherent limitation: it can treat either slow impacts of "not very massive" projectiles, or the final phase of the formation of a crater, in which heating effects have diminished. Future work in this approach could go along two lines: including in more detail the effects of heating, and thus enabling the study of the "hot phase" of the formation of a crater, and exploring the upper mass limits of this approach and introducing (if it turns out to be necessary) some possible new factors which influence the outcome of the process.  


\section{Acknowledgement}
This paper was prepared within the research project $171005$ of the Ministry of Education,Science and Technological Development of Serbia. The author is grateful to Prof. W.Dappen (USC) for great help in finding some of the references. 



\raggedbottom                         

\label{lastpage-01}

\end{document}